\documentstyle[12pt]{article}
\begin{document} 

\title{Black Hole Solution of Quantum Gravity} 
\author{J.\ Kowalski--Glikman\thanks{e-mail 
address 
jurekk@ift.uni.wroc.pl}\\ 
Institute for Theoretical Physics\\ 
University of Wroc\l{}aw\\ 
Pl.\ Maxa Borna 9\\ 
Pl--50-204 Wroc\l{}aw, Poland} 
\maketitle 
 
\begin{abstract} 
We present a spherically symmetric and static exact solution of 
Quantum 
Einstein Equations. This solution is asymptotically (for large $r$) 
identical with the black hole solution on the anti--De Sitter 
background and, for some range of values of the mass possesses two 
horizons. We investigate thermodynamical properties of this solution.
\end{abstract} 
\vspace{12pt} 
PACT number 
04.60 Ds 
\clearpage

\section{Introduction}

In the recent papers \cite{aj} we presented a class of exact 
solutions of
the regularised Wheeler-De Witt equation. In these papers  
we proposed also an interpretation of the resulting
`wave functionals' in terms of the modified gravitational field
dynamics, namely, in the framework the quantum potential approach to 
the 
quantum 
mechanics originally
proposed  by David
Bohm (see e.g., \cite{bohm} and \cite{holland}) and extended to the 
case of
quantum gravity in
\cite{jv} and \cite{sq} (in minisuperspace) and in \cite{jbohm}, 
\cite{shtanov} 
(for
full theory.) The resulting, effective Hamiltonian constraint of 
quantum 
gravity with additional quantum potential terms was presented in 
\cite{aj1}. This Hamiltonian was assumed to be a generator of 
dynamics of 
quantum gravity. In the present paper I present a spherically 
symmetric and 
static solution of such defined theory which can be regarded as a 
black hole 
solution of quantum gravity.

Let us recall the basic steps leading to the quantum Hamiltonian 
constraints. In the papers \cite{aj} two exact solutions of the 
Wheeler--De Witt equation were found:
\begin{eqnarray}
\Psi_I = \exp\left(-\frac{3\rho^{(5)}}{\Lambda}{\cal V}\right) &&
\mbox{Case I};\label{solI}\\
\Psi_{II} = \exp\left( \frac{4\Lambda}{3\kappa^4\rho^{(5)}}{\cal 
R}\right)
&& \mbox{Case II};\label{solII},
\end{eqnarray}
where
${\cal V}=\int \sqrt{h}$ is the volume of the universe,
${\cal R}=\int\sqrt{h}R^{(3)}$ its average curvature, and 
$\rho^{(5)}$ and $\Lambda$ are the renormalisation and bare 
cosmological 
constants. Since the Wheeler--De Witt equation is linear, any complex 
combination of solutions (\ref{solI}, \ref{solII}) is a solution of 
the form 
$\Psi = e^{\Gamma}e^{i\Sigma}$. Taking such a combination, 
substituting to 
the WDW equation, extracting real part, and identifying the 
derivative 
$\frac{\delta S}{\delta h_{ab}}$ with the momenta $p^{ab}$, we get 
the 
equation (in the gauge where the shift vectors $N^a=0$)
$$
0={\cal H}_\bot =  \kappa^2 
G_{abcd}p^{ab}p^{cd} +$$\begin{equation} {\cal F}\left(
\frac{27}{16}\frac{\rho^{(5)}{}^2\kappa^2}{\Lambda^2} \sqrt h
+\frac{1}{\kappa^2} \sqrt h R -
\frac{8}{9} \frac{\Lambda^2}{\kappa^6\rho^{(5)}{}^2}
\sqrt h \left( -\frac38 R^2 + R_{ab}R^{ab}\right)\right), \label{qh}
\end{equation}
\begin{equation}
{\cal F} =
\frac12%
\frac{\sin^2(\phi )}{%
\left\{ \cosh\left(  \frac{3 \rho^{(5)}}{\Lambda} {\cal V} +
\frac{4\Lambda}{3 \kappa^4 \rho^{(5)}}{\cal R}\right)
 + \cos(\phi )
\right\}^{2}},
\end{equation}
where $\kappa$ is the gravitational constant, 
$G_{abcd}$ is the Wheeler--De Witt metric, and $\phi$ is a 
parameter measuring the rate of mixing of two solutions above. 
Generalizing 
the observation of Gerlach \cite{gerlach} (who considered the case 
$\Gamma 
=0$), it is assumed that the quantum Hamiltonian ${\cal H}_\bot$ is a 
generator of time evolution. It should be stressed that the above 
quantum 
Hamiltonian constraint is by no means arbitrary, and can be derived 
from 
exact solutions of quantum gravity given above.

\section{The solution}

In the static case, where momenta are equal to zero, one of the 
dynamical equation of the theory (corresponding to the $00$ component 
of Einstein field equations) is the requirement that the Hamiltonian 
constraint vanishes, to wit
\begin{equation}
\frac{27}{16}\frac{\rho^{(5)}{}^2\kappa^2}{\Lambda^2} \sqrt h
+\frac{1}{\kappa^2} \sqrt h R -
\frac{8}{9} \frac{\Lambda^2}{\kappa^6\rho^{(5)} {}^2}
\sqrt h \left( -\frac38 R^2 + R_{ab}R^{ab}\right) = 0
\label{G00}
\end{equation}
It is worth observing that the cosmological and renormalization 
constants appear only in combination $v_0 = 
\frac{\Lambda}{\rho^{(5)}}$ of 
dimension $m^{3}$. Thus the theory possesses two length scales, the 
Planck 
scale $l$ defined by $\kappa$, and  $v_0^{1/3}$. 

In the spherically symmetric, static case, we can take the three-
metric on 
the hypersurfaces of constant time of the form
\begin{equation}
h_{ab} = A(r) dr^2 + r^2 d\theta^2 + r^2 \sin^2\theta 
d\phi^2.\label{met3}
\end{equation}
Substituting this into equation (\ref{G00}) we find a differential 
equation for the function $A(r)$ 
\begin{eqnarray}
0 &=& A' \left[ \frac89 \frac{v_0^2}{l^6} \frac{A-1}{A^3 r^3} +
\frac{2}{l^2}\frac{1}{A^2 r} \right]\nonumber \\
&+& \left[-\frac49 \frac{v_0^2}{l^6} \frac{(A-1)^2}{A^2 r^4} +
\frac{2}{l^2}\frac{A-1}{A r^2} \right] + \frac{27}{16} 
\frac{l^2}{v_0^2} .\label{HA}
\end{eqnarray}
It is a remarkable fact that the coefficients multiplying $(A')^2$ 
vanishes 
identically. To solve this equation it is convenient to substitute 
$A = (1 - f(r) r^2)^{-1}$. It turns out that the function $f$ 
satisfies the 
quadratic equation
\begin{equation}
f^2 + \frac92\, \frac{l^4}{v_0^2}f + \frac{81}{64}\frac{l^8}{v_0^4} =
\frac94\,\frac{l^6}{v_0^2} \, \frac{\alpha}{r^3},
\label{f}
\end{equation}
where $\alpha$ is an integration constant with the solution
\begin{equation}
f(r)_{\pm} = - \frac94 \frac{l^4}{v_0^2} \pm \frac12\sqrt{ 
\frac{243}{16} 
\frac{l^8}{v_0^4}
+  \frac{9l^6}{v_0^2} \, \frac{\alpha}{r^3}},
\label{fpm}
\end{equation}
and the coefficient $A$ of the three-metric on constant time 
surface equals (we consider only the $f_+$ solution)
\begin{equation}
A = \frac{1}{1+\frac94 \frac{l^4}{v_0^2}r^2 - \frac{ r^2}{2} \sqrt{ 
\frac{243}{16} 
\frac{l^8}{v_0^4}
+ 9 \frac{l^6}{v_0^2} \, \frac{\alpha}{r^3}}}.\label{A}
\end{equation}

In the next step we must construct the full four-metric. This metric 
is 
of the form
$$
g_{\mu\nu} = -N(r)^2dt^2 + A(r) dr^2 + r^2 d\theta^2 + r^2 
\sin^2\theta 
d\phi^2.
$$
To find $N$, one can make use of the dynamical equations presented in 
\cite{aj1} (modified by terms proportional to spacial covariant 
derivatives of $N$), 
but there is a more strightforward way. Namely, we can consider the 
action
\begin{equation}
I =-\frac14 \int N{\cal H}_\bot,\label{I}
\end{equation}
which is the Hamiltonian action for general relativity in the 
static case and in the gauge $N^a=0$. One can easily check that such 
a procedure produces the correct expression for Schwarzschild 
metric. It is convenient to make an anzatz
$$
N = N_0 A^{-1/2}.
$$
Then the action (\ref{I}) takes the form
\begin{equation}
I = - \int N F'dr,\label{IF}
\end{equation}
with 
\begin{equation}
F = \frac{v_0^2}{9l^6}\, \frac{(A-1)^2}{A^2 r} +
\frac{1}{2 l^2}\, r \, \frac{A-1}{A} + \frac{9 l^2}{64 v_0^2}\, r^3.
\end{equation}
The expression from $A$, (\ref{A}), as well as the condition $N_0 = 
const$ 
follow from (\ref{IF}).

Thus our solution is of the form (we take $N_0 =1$ for a while)
$$
g_{\mu\nu} =
- \left(
1+\frac94 \frac{l^4}{v_0^2}r^2 - \frac{r^2}{2} \sqrt{ \frac{243}{16} 
\frac{l^8}{v_0^4}
+ 9 \frac{l^6}{v_0^2} \, \frac{\alpha}{r^3}}\right) dt^2 +$$
\begin{equation} +
\frac{1}{1+\frac94 \frac{l^4}{v_0^2}r^2 - \frac{r^2}{2} \sqrt{ 
\frac{243}{16} 
\frac{l^8}{v_0^4}
+ 9 \frac{l^6}{v_0^2} \, \frac{\alpha}{r^3}}} dr^2 + r^2 (d \theta +
\sin^2\theta d\phi^2) \label{met4}
\end{equation}
\vspace{12pt}

Let us now analyse the solution. First consider the asymptotics for 
large 
$r$:
\begin{equation}
A(r) \stackrel{r \rightarrow \infty}{\sim} \left(1+ \frac{r^2}{8} 
\frac{l^4}{ v_0^2}(18-\sqrt{243}) - \frac{l^2}{\sqrt 3 } 
\frac{\alpha}{r}\right)^{-1}.
\end{equation}
We see therefore that our solution is asymptotically equivalent to 
the black hole solution on the anti--De Sitter background with the 
effective cosmological constant 
$\lambda = -\frac{l^4}{8 v_0^2}(18-\sqrt{243})$. Thus we interpret 
$\alpha$ 
as being proportional to the black hole mass $M$, so that the last 
term 
becomes the familiar $\frac{2 l^2 M}{r}$. This means that
$$
\alpha = 2\sqrt 3 M 
$$

The asymptotics at $r$ tending to zero is regular
\begin{equation}
A(r) \stackrel{r \rightarrow 0}{\sim} 
\left(1-\frac{2 l^3 }{3 v_0}\sqrt{2\sqrt{3} M r} \,\,\right)^{-1}.
\end{equation}
In spite of that the invariants constructed from the Riemann tensor 
are divergent at $r = 0$. The singularity is of  order 
(Riemann)${}^2$ $\sim r^{-3}$ and is significantly milder than the 
singularity of Schwarzschild solution, where it is of order $\sim r^{-
6}$. This singularity is hard to avoid because the solution describes 
a point mass source. 

Thus we have the four metric $g_{\mu\nu}$ (\ref{met4}) that 
describes  an exact static, spherically symmetric black hole solution 
of 
quantum gravity (with negative cosmological constant).  
\newline

For large $M$ the space--time described by (\ref{met4}) has two 
horizons 
at $r_+$ and $r_-$, whose exact values can be found by solving 
quartic 
equation.  When $M$ decreases, the horizons come closer to each 
other, and 
finally they degenerate to a single one at $r_c$. For still smaller 
values 
of $M$ there are no horizons. This behaviour is similar to the way 
the 
standard charged black hole solution behaves, when its mass 
decreases, 
aproaching the charge from above.

It can be checked that the function $A(r)^{-1}$ has only one minimum 
for 
positive $r$. This means that at the point where horizons merge (for 
some 
specific $M$), the derivative $\left.\left(\frac{d}{dr}\, A(r)^{-
1}\right) 
\right|_{r_c} =0$. This is important, because, as we will see, it 
follows 
that the temperature of the extreme black hole vanishes.
 
One can check that the  causal structure of our solution is similar 
to 
that of the Reissner--Nordstr\"om solution on the anti -- De Sitter 
background; the corresponding Penrose diagram can be found, for 
example, in 
\cite{Teit}.

\section{Thermodynamics}

One can analyse the thermodynamics of quantum black hole basically 
repeating 
the standard steps \cite{GH} (see also \cite{Teit}). To find the 
temperature 
we impose regularity on the horizon of the Euclidean continuation of 
the 
metric (\ref{met4}), $t \rightarrow -i\tau$ defined for $r > r_+$ 
\begin{equation}
ds_E^2 = N_0 A^{-1}(r) d\tau ^2 + A(r) dr^2 + r^2 d 
\Omega^2,
\end{equation}
with $A(r)$ given by (\ref{A}). To avoid conical 
singularity on the horizon $r=r_+$, we take $\tau $ periodic with the 
range 
$[0,1]$. Then we can identify $N_0$ with the inverse temperature 
$N_0= 
T^{-1}$, which can be found from standard formula
\begin{equation}
T = 
\frac{1}{4\pi } \left(\frac{\partial A^{-1}(r)}{\partial 
r}\right)_{r=r_+} 
.\label{temp}
\end{equation}

As mentioned already, for critical mass, where two horizons merge and 
form 
the single critical one, the derivative of $A^{-1}$ vanishes. Thus 
the 
extreme black hole has zero temperature.

Let us try to rewrite the above expression for temperature in a more 
explicit form. The equation for horizon reads
$$
1 - r_+ f(r_+) =0.
$$
Differentiatin equation (\ref{f}) with respect to $r$ and putting $r 
= r_+$ 
we find
$$
f'(r_+) = - \frac{3 \alpha}{2 r_+^2}\, \frac{1}{r_+^2 + \frac{4 
v_0^2}{9 
l^4}}
$$
Moreover, it follows from (\ref{f}) that
\begin{equation}
\alpha = \frac{4v_0^2}{9 l^6}\, \frac{1}{r_+} + \frac{2}{l^2} r_+
+\frac{9 l^2}{16 v_0^2} r_+^3.
\end{equation}
The final expression for the temperature reads
\begin{equation}
T = \frac{1}{4\pi} \left\{ - \frac{2}{r_+}
+ \frac{3 }{2 }\, \frac{1}{r_+^2 + \frac{4 v_0^2}{9 
l^4}} \left( \frac{4v_0^2}{9 l^6}\, \frac{1}{r_+} + \frac{2}{l^2} r_+
+\frac{9 l^2}{16 v_0^2} r_+^3 \right) \right\}.
\end{equation}
For large masses i.e., large $r_+$ we have
\begin{equation}
T \sim \frac{1}{4\pi}\left[ \frac{27 l^2}{32 v_0^2} r_+ + \frac58 \,
\frac{1}{r_+}\right].
\end{equation}
This last expression is up to numerical coefficients identical with 
the 
classical result for black hole on anti -- De Sitter background 
\cite{Teit}.
\newline

In the next step we can compute the entropy of the solution. To this 
end we 
use the procedure proposed by Gibbons and Hawking \cite{GH}, and we 
identify 
the action $I_E$ (\ref{IF}) with free energy 
divided by temperature:
\begin{equation}
I_E = \frac{M}{T} -S.
\end{equation}
However, as it is well known we must add boundary terms $B$ at 
infinity 
and at $r = r_+$ which are fixed by boundary conditions of the 
variational 
problem for the action $I_E$. Thus we consider
$$
I_E = -\int_{r_+}^{\infty} N_0 F' + B
$$
The integral is a linear combination of constraints ${\cal 
H}_\bot(r)$, and 
thus
\begin{equation}
B =  \frac{M}{T} -S.
\end{equation}
The variation of the action up to the terms that vanish on-shell 
equals
\begin{equation}
\delta I_E =
- \left[ N \delta F \right]_{r_+}^{\infty} + \delta B(\infty) + 
\delta 
B(r_+).
\end{equation}
Consider the variation at $r_+$. Using the condition
\begin{equation}
4 \pi = 
N(r_+) \left(\frac{\partial A^{-1}(r)}{\partial r}\right)_{r=r_+} 
\end{equation}
we get the condition
\begin{equation}
\delta B(r_+) = 4 \pi \left(\frac{\partial F}{\partial 
A^{-1}(r)}\right)_{r=r_+} .
\end{equation}
Using the condition $N(\infty) = 1/T$ and solving the boundary 
condition at 
infinity we find
\begin{equation}
B = \frac{M}{T} + 4 \pi \int\, dr_+ \left(\frac{\partial F}{\partial 
A^{-1}(r)}\right)_{r=r_+} - S_0,
\end{equation}
where $S_0$ is a constant to be fixed in a moment.
The entropy of the black hole of outer radious $r_+$ is therefore 
equal
$$
S = -4 \pi \int\, dr_+ \left(\frac{\partial F}{\partial 
A^{-1}(r)}\right)_{r=r_+} + S_0 =
$$
$$
=\frac{8\pi v_0^2}{l^6} \log(r_+) + \frac{\pi}{l^2} r_+^2 +S_0
$$
Assuming that the entropy vanishes for extreme black hole with $r_+ = 
r_c$, 
we finally get
\begin{equation}
S = \frac{8\pi v_0^2}{l^6} \log\left(\frac{r_+}{r_c}\right) + 
\frac{\pi}{l^2} (r_+^2 - r_c^2)
\end{equation}
For large $r_+$ the entropy is thus given by the standard Bekenstein -
- 
Hawking formula, and is equal to $1/4$ of the area of the black hole. 
There 
is another logarythmic term, whose interpretation is not clear.

\section{Conclusions}

\begin{enumerate}
\item The reader may wonder why we have chosen solution $f_+$ in 
(\ref{f}). 
The reason is that if solution $f_-$ was to represent a black hole 
with 
positive mass, then the coefficient $\alpha$ would have to be 
negative, and 
the space-time would develop additional circular singularity (for $r$ 
such 
that the expression in squire root in (\ref{A}) vanishes. If $\alpha$ 
is 
positive we have to do with a negative mass black hole, whose 
interpretation 
is not clear.

\item Of course the most important problem to be analized is the 
issue of 
Hawking radiation. To approach this problem one should in principle 
has in 
disposal a quantum theory of gravitational field coupled to some 
matter 
field. Solutions of quantum gravity coupled to the massles scalar 
field 
exist \cite{sc} and it is feasible to construct a quantum hamiltonian 
in 
this case. It is clear however that if the quantum black hole does 
radiate 
and its mass decreases, the temperature tends to zero. Contrary to 
the 
standard black hole, where the evaporation process becomes more and 
more 
rapid as the mass decreases $T \sim M^{-1}$, here the process 
gradually stops
and the black hole leaves eventually a cold remnant of the size of 
order of 
Planck length (assuming that the ratio $l^3/v_0$ is of order $1$.) 
These 
questions are currently under investigation.
\end{enumerate}

{\bf Acknowledgment}

I would like to thank professor M.\ Demianski for discussion, and  
A.\ 
B\l{}aut and D.\ Nowak for their help in checking some part of 
calculations.


\begin{thebibliography}{99} 

\bibitem{aj} J.\ Kowalski-Glikman and K.\ Meissner, Phys.\ Lett.\
{\bf  B 376} (1996) 48;
A.\ B\l{}aut and J.\ Kowalski-Glikman 
{\em Constraints nad Solutions of Quantum Gravity in Metric 
Representation},
gr-qc 9710037.
\bibitem{bohm}  D.\ Bohm, Phys.\ Rev.\ {\bf 85}, {166} (1952), 
Phys.\ Rev.\ {\bf 85}, 180 (1952); 
D. Bohm and B.J.\ Hiley, Phys.\ Rep.\ {\bf 144}, 323 (1987); 
D. Bohm, B.J.\ Hiley, and P.N.\ Koloyerou Phys.\ Rep.\ {\bf 144}, 349 
(1987); 
D. Bohm, B.J.\ Hiley, {\em The Undivided Universe: An Ontological 
Interpretation of Quantum Theory}, Routledge \& Kegan Paul, London, 
1993; 
J.\ S.\ Bell {\em Speakable and Unspeakable in Quantum 
Mechanics}, Cambridge University Press, 1987.
\bibitem{holland} P.R.\ Holland, 
{\em The
Quantum Theory of Motion}, Cambridge University Press, 1993.
\bibitem{jv} J.\ Kowalski-Glikman and J.\ Vink, Class.\ Quantum Grav.,
{\bf 7}, 901 (1990); J.\ Vink, Nucl.\ Phys. {\bf B 369}, 707 (1992).
\bibitem{sq} E.\ Squires, Phys.\ Lett.\
{\bf  A 155} (1991) 357.
\bibitem{jbohm} J.\ Kowalski-Glikman; {\em Quantum Potential and 
Quantum 
Gravity} in From Field Theory to Quantum Groups, B.\ Jancewicz 
and J.\ Sobczyk
eds., World Scientific, 1996;
gr-qc 9511014. 
\bibitem{shtanov} Yu.\ Shtanov, Phys.\ Rev.\ {\bf D 54}, 
2564 (1996).
\bibitem{gerlach} Ulrich H.\ Gerlach, Phys.\ Rev.\ {\bf 177}, 
1929 (1969).
\bibitem{aj1} A.\ B\l{}aut and J.\ Kowalski-Glikman 
{\em The Time Evolution of Quantum
Universes in the Quantum Potential Picture},
gr-qc 9710136.
\bibitem{Teit} M\'aximo Ban\~ados, Claudio Teitelboim, Jorge Zanelli, 
Phys.\ Rev.\ {\bf D 49}, 975 (1994).
\bibitem{GH} G.\ Gibbons and S.W.\ Hawking, Phys.\ Rev.\ {\bf D 15}, 
2753 (1977).
\bibitem{sc} A.\ B\l{}aut and J.\ Kowalski-Glikman, Phys.\ Lett.\ 
{\bf B 
406}, 33 (1977).



\end{thebibliography}
\end{document}